\documentclass[a4paper,11pt]{article}
\usepackage{pos}
\usepackage{amsmath}
\usepackage{graphicx}
\usepackage{float}
\usepackage{soul}
\usepackage[export]{adjustbox}
\usepackage[utf8]{inputenc}
\usepackage{slashed}
\usepackage{subfig}
\usepackage{scalerel}
\usepackage{cancel}
\usepackage{subcaption}
\usepackage{pifont}
\usepackage{xcolor}
\usepackage{physics}
\usepackage{enumitem}
\usepackage{ifthen}
\usepackage{mathtools}
\usepackage{import}
\numberwithin{equation}{section}
\usepackage{standalone}
\usepackage{tikz}
\usepackage{comment}
\usepackage[compat=1.1.0]{tikz-feynman}
\usetikzlibrary{decorations.markings,arrows,arrows.meta,shapes}
\usetikzlibrary{decorations.text}
\usetikzlibrary{bending} 
\DeclareUnicodeCharacter{2212}{-}

\renewcommand{\logo}{\relax}

\title{Ward identity preserving local ultraviolet counterterms for photoproduction at two loops in QCD}
\ShortTitle{Local UV counterterms for photoproduction in QCD}
\author[a]{Charalampos Anastasiou}
\author[a]{Julia Karlen}
\author*[a]{Roshni Sahoo}
\author[b]{George Sterman}
\author[a]{Matilde Vicini}

\affiliation[a]{Institute for Theoretical Physics, ETH Zurich,\\8093 Z\"urich, Switzerland}
\affiliation[b]{C.N.\ Yang Institute for Theoretical Physics and Department of Physics and Astronomy, Stony Brook University,\\Stony Brook NY, 11794-3840 USA}

\emailAdd{babis@phys.ethz.ch}
\emailAdd{karlenj@phys.ethz.ch}
\emailAdd{rsahoo@phys.ethz.ch}
\emailAdd{mvicini@phys.ethz.ch}
\emailAdd{george.sterman@stonybrook.edu}

\abstract{ We review the construction of locally finite two-loop amplitude integrands for photoproduction via quark annihilation, presented in Ref.\,\cite{Anastasiou:2025cvy}. Building on established techniques for off-shell colorless production, we extend this local subtraction framework to handle transient singularities arising from real outgoing photons. These singularities manifest only at the integrand level, yielding finite contributions upon integration. In these proceedings, we provide the explicit construction of ultraviolet counterterms that satisfy necessary Ward identity cancellations, ensuring that the integrand is rendered integrable. This work provides a locally finite amplitude integrand that is ready for numerical integration in momentum space. Furthermore, it establishes the foundation for extending local subtraction frameworks to processes involving final-state jets.} 

\FullConference{17th International Symposium on Radiative Corrections: Applications of Quantum Field Theory to Phenomenology (RADCOR2025)\\
5-10 October 2025\\
Puri, India\\}

\allowdisplaybreaks

\begin{document}

\renewcommand{\hookAfterAbstract}{%
\par\bigskip

}
\maketitle

\section{Introduction}
The development of a process-independent algorithm for the numerical integration of multi-loop amplitudes in momentum space relies on the construction of locally finite integrands. Recent progress in this framework has established techniques for constructing locally finite two-loop amplitudes for several processes including off-shell colorless production via quark annihilation \cite{Anastasiou:2022eym,Anastasiou:2020sdt,Anastasiou:2024xvk}. Most recently, this approach has been extended to real-virtual corrections to cross sections at NNLO for colorless hadroproduction \cite{Anastasiou:2026kpm}. This work extends the framework in Ref.\,\cite{Anastasiou:2022eym} to real final-state photons \cite{Anastasiou:2025cvy}.

Real photons introduce \textit{transient} infrared (IR) singularities that manifest at the integrand-level. These singularities must be eliminated from the integrand to enable numerical integration. We present methods to eliminate these singularities using local counterterms that integrate to zero, and preserve initial-state IR factorisation locally. Complementing the description of our work in \cite{Anastasiou:2025cvy}, these proceedings present the construction of local ultraviolet (UV) counterterms consistent with local IR factorisation. Section \ref{review} reviews the subtraction of transient IR singularities, followed by the treatment of UV singularities in Section \ref{UV}.  

\section{Transient Infrared Singularities} \label{review}
 We present the removal of final-state transient singularities using diphoton production $ q(p_{1}) + \bar{q}(p_{2}) \rightarrow \gamma(q_{1}) + \gamma(q_{2}) $, though our methods generalize to any multiplicity of colorless final states. Transient singularities appear in two-loop diagrams that have either a one-loop correction at the outgoing real photon vertex or a nested self-energy correction on the adjacent quark propagator.  
\subsection{Nested self-energy corrections}
We consider the subleading color ($-C_{F}/2N$) part of the diagrams in Eq.~\eqref{eq:SLCgroupSE}, denoted as ${\mathcal A}^{(2), {\rm SLC}}(k,l)$. These diagrams locally factorise in all initial-state soft and collinear limits ($k \parallel p_{1}$ or $k \parallel p_{2}$).  
\begin{align}
\label{eq:SLCgroupSE}
& {\cal A}^{(2), {\rm SLC}}_{q\bar q \to \gamma \gamma}(k,l) 
=  -\frac{1}{2 \, N \,C_F} 
\left( \includegraphics[scale=0.55, page=1, valign=c]{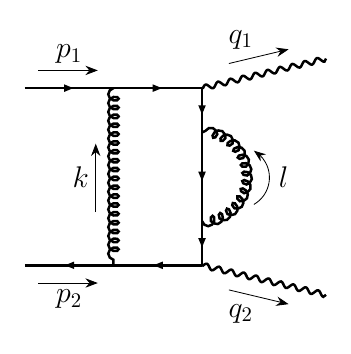}
    + \includegraphics[scale=0.55, page=3, valign=c]{figures/self-energy.pdf}
    + \includegraphics[scale=0.55, page=6, valign=c]{figures/self-energy.pdf}
    \right. 
\nonumber \\
& \left.  
+ \includegraphics[scale=0.5, page=8, valign=c]{figures/self-energy.pdf}
+ \includegraphics[scale=0.5, page=9, valign=c]{figures/self-energy.pdf}
    \right) 
    + \includegraphics[scale=0.5, page=10, valign=c]{figures/self-energy.pdf}
    + \includegraphics[scale=0.5, page=11, valign=c]{figures/self-energy.pdf} \, 
\end{align}
Diagrams with nested self-energy subgraphs, such as,
\begin{eqnarray}
\label{eq:rRSR-integrand}
\includegraphics[scale=0.55, page=5, valign=c]{figures/self-energy.pdf}
= \ldots  
\frac{i}{\slashed R}  \, 
{\cal S}(l, R) \, 
\frac{i}{\slashed R}  \, 
\slashed \epsilon_1^* \,   
\frac{i}{\slashed r } 
\ldots \,,  \quad r = k+p_1 \,,  R= k+p_1 -q_1 \,, 
\nonumber\\
\end{eqnarray}
contain doubled propagators $i/\slashed{R}$, which result in power-like singularities of the integrand as $r = k+ p_{1}$ becomes parallel to the outgoing real photon $q_{1}$. To eliminate these singularities, one repeated propagator must be removed via a tensor reduction of the self-energy subgraph. This reduction is achieved by averaging the sub-graph over the equivalent loop momentum flows $l$ and $-l-R$. 

To preserve the Ward identity cancellations required for the factorisation of initial-state singularities, this symmetrization must be applied to the entire set in Eq.~\eqref{eq:SLCgroupSE}, 
\begin{eqnarray}
\label{eq:SLCsymmSErecipe}
 {\mathcal A}^{(2), {\rm SLC}}(k,l) 
 &\to& \langle {\cal A}^{(2), {\rm SLC}}_{\rm IR}(k,l) \rangle_{l} \nonumber \\ 
 &=& \frac{1}{2}
 {\cal A}^{(2), {\rm SLC}}(k,l)
 + \frac{1}{2}{\cal A}^{(2), {\rm SLC}}(k,-l-R), 
\end{eqnarray}
where $\langle \ldots \rangle_{l}$ denotes the symmetrization of the integrand under the exchange $ l \leftrightarrow -l -R$. 

In the initial-state collinear limits, $\langle {\cal A}^{(2), {\rm SLC}}(k,l) \rangle_{l}$ factorizes into a one-loop form factor times a hard tree-level integrand. This structure is captured and removed by the IR counterterm $\langle {\cal A}^{(2), {\rm SLC}}_{IR}(k,l) \rangle_{l}$,
\begin{eqnarray} \label{eq:SLC_IR_shorthand}
\langle {\cal A}^{(2), {\rm SLC}}(k,l) \rangle_{l}
\sim \langle {\cal A}^{(2), {\rm SLC}}_{\rm IR}(k,l) \rangle_{l}
\end{eqnarray}
where 
\begin{align}
    \label{eq:SLC_IR_approx}
    {\cal A}^{(2), {\rm SLC}}_{\rm IR}(k,l)  
    = \frac{-1}{2 N C_F} 
    \vcenter{\hbox{\includegraphics[scale=0.5, page=1]{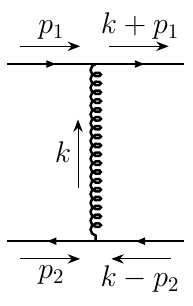}}} \times \, 
    \left( \vcenter{\hbox{\includegraphics[scale=0.5, page=2]{figures/radcor-diagrams.pdf}}}  \right) , \quad  \mathbf{P} = \frac{\slashed{p}_{1}\slashed{p}_{2}}{2 p_{1} \cdot p_{2}}, 
\end{align}
rendering the integrand locally finite in the initial and final-state singular regions. This loop momentum symmetrization is analogously performed for diagrams with leading color, as described in Ref.\,\cite{Anastasiou:2025cvy}. 

\subsection{Nested triangle subgraphs}
\begin{figure}[h]
    \centering
    \includegraphics[scale=0.5,valign=c, page=4]{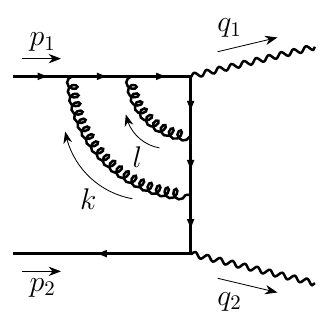}
    \qquad
    \includegraphics[scale=0.5,valign=c, page=9]{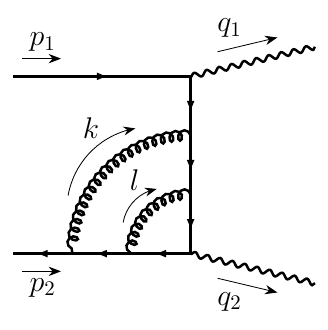}
    \qquad
    \includegraphics[scale=0.5,valign=c, page=7]{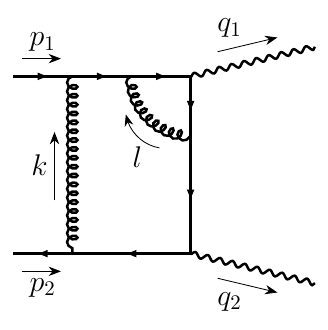}
    \qquad
    \includegraphics[scale=0.5,valign=c, page=13]{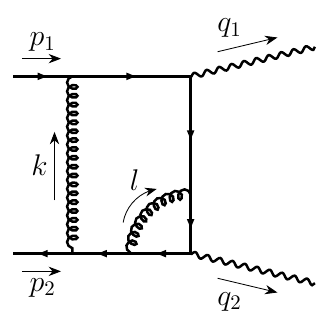}
    \caption{Two-loop diagrams with one-loop  corrections to real photon vertices. }
    \label{fig:planQEDtriangle}
\end{figure}

A second class of transient singularities arises from diagrams with one-loop corrections to the outgoing photon vertex, shown in Fig.\ref{fig:planQEDtriangle}. Unlike the one-loop case, the numerator does not vanish in the limits $k + p_{1} \parallel q_{1}$ and $k - p_{2} \parallel q_{2}$.  
To remove these transient singularities, we proposed a solution in  Ref.\,\cite{Anastasiou:2025cvy} involving loop momentum symmetrization in the transverse plane relative to the photon momentum and an auxiliary light-like vector. Alternatively, these singularities can be addressed via a tensor reduction of the divergent integrand components. These complementary methods, offering alternative solutions via transverse symmetrization or tensor reduction, were recently demonstrated for initial-state loop polarizations in Ref.~\cite{Anastasiou:2026kpm}. In these proceedings, we present the counterterm derived from the tensor reduction solution, which integrates to zero and  cancels the divergence at the integrand level,  
\begin{align} \label{eq:transient-limit}
&\frac{N}{2 C_F} 
\includegraphics[scale=0.4,valign=c, page=3]{figures/loop-pol.pdf}   
-\frac{1}{2 N C_F} \includegraphics[scale=0.4,valign=c, page=1]{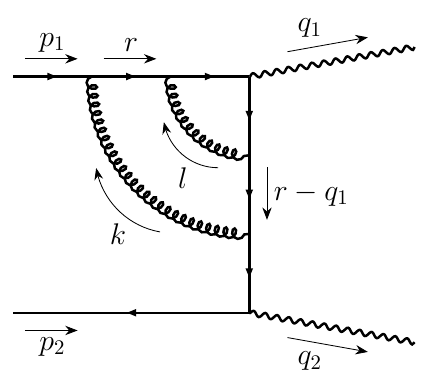} + \includegraphics[scale=0.4,valign=c, page=32]{figures/UV-diags.pdf} 
&\xrightarrow[]{r=xq_1} 0 \, ,
\end{align}
where the vertex $\Delta^{\mu}_{\gamma}(r,q,k,l)$ is defined as
\begin{align}  \label{eq:DeltaTransient}
    & \Delta^{\mu}_{\gamma}(r,q,k,l) = g_{s}^{2} \, e \, (2-D) \, C_{F}\, \frac{\slashed{\eta}}{q \cdot \eta} \left[ \frac{N}{2C_{F}} (r+2(l-k))^{\mu} \, \frac{(r + l -k)\cdot q }{(l-k)^{2} (l-k+r-q)^{2} (l -k + r)^{2}} \right. \nonumber \\
    & \hspace{4cm}\left. -\frac{1}{2NC_{F}} (r + 2l)^{\mu} \frac{(r+l)\cdot q}{l^{2} (l + r - q)^{2} (l+r)^{2}} \right] . 
\end{align}
The counterterm vertex in Eq.~\eqref{eq:DeltaTransient} applies to all diagrams containing one-loop photon vertex corrections. Its construction relies on decomposing the loop momenta ($r + l$ for subleading and $r + l - k$ for leading color) into components parallel to the photon momentum $q$, an auxiliary light-like reference vector $\eta$ and the transverse plane. For any vector $v$, this decomposition is defined as,
\begin{eqnarray}
\label{eq:v-project}
v^\mu &=& \frac{v \cdot \eta}{q\cdot \eta} q^\mu
+  \frac{v \cdot q}{q\cdot \eta} \eta^\mu + v_{\perp(q,\eta)}^\mu\, , 
\end{eqnarray}
where we require $ q \cdot \eta \neq 0$. 
The treatment of initial-state IR divergences in the $k\parallel p_{1}$ or $p_{2}$ limits, arising in the counterterms with $\Delta_{\gamma}^{\mu}$ vertex, is described in detail in Ref.\,\cite{Anastasiou:2025cvy}.  
\newpage 
\section{Ultraviolet Singularities} \label{UV}
To enable numerical integration of two-loop amplitudes in 4 dimensions, the integrand must be locally finite in the ultraviolet (UV) region. This requires local counterterms for divergent configurations where any combination of loop momenta become large. For single-loop sub-divergences, we construct subtractions for every divergent subgraph, such as self-energy and one-loop vertex corrections, via a Taylor expansion in the large momentum limit, truncating the series to retain only the divergent terms. The complete UV subtraction is organized using the forest formula \cite{Bogoliubov:1957gp, Hepp:1966eg,  Zimmermann:1969jj,Caswell:1981ek, Capatti:2022tit}
, while in the following sections we focus on the construction of UV counterterms for one loop subgraphs. This allows us to demonstrate the Ward identity cancellations required to preserve local IR factorisation in the presence of UV subtractions.
\subsection{Nested self-energy corrections}
The self-energy and one-loop vertex correction sub-graphs in $\mathcal{A}^{(2),SLC} (k,l)$ exhibit UV divergences as $l \to \infty$. Simultaneously, they exhibit IR singularities as the secondary loop momentum $k$ approaches soft or collinear limits. The UV counterterms in $l$ must be constructed to preserve the Ward identity cancellations that enable local factorisation in the soft and collinear limits of $k$. Specifically, the UV expanded integrand must satisfy the IR approximation in Eq.~\eqref{eq:SLC_IR_shorthand} for large $l$. 
To illustrate this construction, consider the self-energy subgraph
\begin{align}
    \includegraphics[scale=0.6, page=17, valign=c]{figures/UV-diags.pdf} = -\,g_{s}^{2}\,C_{F}\,\gamma_{\mu}\frac{(\slashed{l}+\slashed{P} + \slashed{Q})}{(l+Q)^{2}(l+P+Q)^{2}} \gamma^{\mu}, 
\end{align}
where the momenta $P$ and $Q$ are fixed scales (linear combinations of $\{k,p_{i},q_{i} \}$) relative to the large loop momentum $l$. In the diagram above, the propagators marked with a dot are truncated, indicating its exclusion from the integrand expression on the right hand side. The UV counterterm for the self-energy correction in the limit $l \to \infty$ is given by, 
\begin{align} \label{eq:UV-operator}
    \includegraphics[scale=0.6, page=17, valign=c]{figures/UV-diags.pdf} \xrightarrow{l \to \infty} -\,g_{s}^{2}\,C_{F}\,(2-D)\,\left( \frac{\slashed{l}+\slashed{P}+\slashed{Q}}{(l^{2} - M^{2})^{2}} - 2 \frac{ l\cdot (P + 2Q) \, \slashed{l}}{(l^{2} - M^{2})^{3}} \right) \equiv \includegraphics[scale=0.6, page=10, valign=c]{figures/UV-diags.pdf}, 
\end{align}
where the \textit{cross} operator defined in Eq.~\eqref{eq:UV-operator} denotes the UV counterterm for any self-energy graph. Additionally, the mass $M$ in the denominators serves as an infrared regulator. The abelian one-loop vertex UV counterterm is constructed similarly, \pagebreak 
\begin{align} \label{eq:abelian-vertexUV}
    \includegraphics[scale=0.6, page=3, valign=c]{figures/UV-diags.pdf} \xrightarrow[]{l \to \infty} &\includegraphics[scale=0.6, page=11, valign=c]{figures/UV-diags.pdf} \nonumber  \\
    \equiv &-g_{s}^{3} \left( - \frac{1}{2N}T^{c}\right) \, (2-D) \, \left[ \frac{2l^{\mu} \slashed{l}}{(l^{2} - M^{2})^{3}}  - \frac{\gamma^{\mu}}{(l^{2} - M^{2})^{2}}\right] . 
\end{align}
The above counterterm is constructed such that the total integrand is locally factorised in the infrared regions. To demonstrate this, consider the $k \parallel p_{1}$ limit of the UV counterterms $\mathcal{A}^{(2),SLC}_{UV} (k,l)$ for the subleading color diagrams defined in Eq.~\eqref{eq:SLCgroupSE}.  
\begin{align} \label{eq:SLC-UV-IRlimit}
    &\mathcal{A}^{(2),SLC}_{UV} (k ,l) \nonumber \\ 
    &\xrightarrow[]{k \parallel p_{1}} -\frac{1}{2NC_{F}} \left( \includegraphics[scale=0.5, page=13, valign=c]{figures/UV-diags.pdf} + \includegraphics[scale=0.5, page=14, valign=c]{figures/UV-diags.pdf} + \includegraphics[scale=0.5, page=16, valign=c]{figures/UV-diags.pdf} \right) \nonumber \\ 
    & \hspace{2cm}+ \includegraphics[scale=0.5, page=15, valign=c]{figures/UV-diags.pdf} . 
\end{align}
In the $k\parallel p_{1}$ limit, the gluon carrying momentum $k$ becomes longitudinal, indicated graphically by an arrow. This allows the application of the Ward identity to the quark-gluon vertex, 
\begin{align}
   k_{\mu} \, \frac{i}{\slashed{p}-\slashed{k}}\, \left( -ig_{s} \gamma^{\mu} T^{c}\right) \, \frac{i}{\slashed{p}}  = g_{s} T^{c} \left( \frac{i}{\slashed{p} - \slashed{k}} - \frac{i}{\slashed{p}}\right ) . 
\end{align}
Graphically,  
\begin{align} \label{eq:ward-graphical}
    \includegraphics[scale=0.6, page=6, valign=c]{figures/radcor-diagrams.pdf} =  \left( \includegraphics[scale=0.6, page=9, valign=c]{figures/radcor-diagrams.pdf} - \includegraphics[scale=0.6, page=10, valign=c]{figures/radcor-diagrams.pdf}\right) ,
\end{align}
where the cross symbol denotes the absorption of the factor $g_{s}T^{c}$ and implies a truncated propagator. 
\pagebreak
By applying the Ward identity in Eq.~\eqref{eq:ward-graphical} to the diagrams in Eq.~\eqref{eq:SLC-UV-IRlimit}, we obtain 
\begin{align} \label{eq:SLC-after-ward}
    &\mathcal{A}^{(2),SLC}_{UV} (k ,l)  \nonumber \\ 
    &\xrightarrow{k \parallel p_{1}} -\frac{1}{2NC_{F}}  \left( \includegraphics[scale=0.5, page=18, valign=c]{figures/UV-diags.pdf} + \includegraphics[scale=0.5, page=19, valign=c]{figures/UV-diags.pdf} - \includegraphics[scale=0.5, page=20, valign=c]{figures/UV-diags.pdf} \nonumber \right. \\ 
    & \left.+ \includegraphics[scale=0.5, page=22, valign=c]{figures/UV-diags.pdf} - \includegraphics[scale=0.5, page=21, valign=c]{figures/UV-diags.pdf} \right) +  \includegraphics[scale=0.5, page=15, valign=c]{figures/UV-diags.pdf}  
\end{align}
In Eq.~\eqref{eq:SLC-after-ward}, the first and the third diagrams cancel as they are identical with opposite signs. The remaining terms (2, 5 and 6) undergo a cancellation through the following identity relating the self-energy and vertex correction UV counterterms,   
\begin{align} \label{eq:QED-UV-Ward}
    &-\frac{1}{2NC_{F}}\,  \left( \includegraphics[scale=0.5, page=24, valign=c]{figures/UV-diags.pdf} - \includegraphics[scale=0.5, page=23, valign=c]{figures/UV-diags.pdf}\right) \nonumber \\
    &= -g_s^{3} \left( -\frac{1}{2N} T^{c}\right) (2-D) \left[  \frac{2(l\cdot k) \slashed{l}}{(l^{2}-M^{2})^{3}} - \frac{\slashed{k}}{(l^{2}-M^{2})^{2}}\right ]  
     = \, \includegraphics[scale=0.5, page=12, valign=c]{figures/UV-diags.pdf} .   
\end{align}
This yields the UV approximation of the IR counterterm for $\mathcal{A}^{(2),SLC}(k,l)$ in the $k \parallel p_{1}$ limit, 
\begin{align}
    \mathcal{A}^{(2),SLC}_{UV} (k ,l)   
    \xrightarrow{k \parallel p_{1}} -\frac{1}{2NC_{F}} \includegraphics[scale=0.5, page=22, valign=c]{figures/UV-diags.pdf} . 
\end{align}
While the Ward identity for the UV counterterms was demonstrated in Eq.~\eqref{eq:QED-UV-Ward} for the momentum routing $l$, it remains valid for the shifted routing $-l-R$ in $\mathcal{A}^{(2),SLC}(k,-l-R)$. 
This ensures that local factorisation of the symmetrized amplitude in Eq.~\eqref{eq:SLC_IR_shorthand}, is preserved in the UV expansion. 

For leading color contributions, we define the UV counterterm for the non-abelian vertex correction, which arises from the scalar decomposition of the triple-gluon vertex (Eq.~(3.10) in Ref.\,\cite{Anastasiou:2025cvy}), 
\begin{align}
    \includegraphics[scale=0.5, page=7, valign=c]{figures/UV-diags.pdf} \xrightarrow{l \to \infty} 
    g_{s}^{3} \left( \frac{N}{2} T^{c} \right) (2-D) \frac{2l^{\mu} \slashed{l}}{(l^{2} -M^{2})^{3}} . 
\end{align} 
Parallel to the subleading color, these leading color UV counterterms also satisfy the Ward identity like in Eq.~\eqref{eq:QED-UV-Ward}. 

To ensure that the integrand is locally finite in all the IR singular regions, the integrand symmetrization performed to remove transient singularities emerging from nested self-energies must be extended to the UV counterterms. This symmetrization achieves a tensor reduction that effectively removes the doubled propagator $i/\slashed{R}$. In the $l \to \infty$ limit, the averaged self-energy evaluates to, 
\begin{align}
    \frac{1}{2} \includegraphics[scale=0.5, page=1, valign=c]{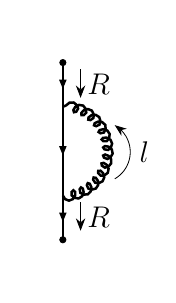} + \frac{1}{2 } \includegraphics[scale=0.5, page=2, valign=c]{figures/UV-diags.pdf} &\xrightarrow[]{l \to \infty } 
     - \frac{1}{2} g_{s}^{2} C_{F} (2-D) \frac{1}{(l^{2}-M^{2})^{2}} \slashed{R} . 
\end{align}
The resulting expression is independent of $l$, and upon insertion into the full diagram, it eliminates the doubled propagator, rendering the UV subtracted integrand locally free of the transient singularities. 

\subsection{Nested triangle subgraphs}
For the triangle subgraphs in Fig.\,\ref{fig:planQEDtriangle}, the UV counterterms must preserve the cancellation of the transient singularities in the $k + p_{1} \parallel q_{1}$ and $k-p_{2} \parallel q_{2}$ limits. In the UV limit $l \to \infty$, the triangle vertex correction is given by,
\begin{align}\label{eq:abelian-vertex-UV2}
    \includegraphics[scale=0.5, page=29, valign=c]{figures/UV-diags.pdf} \xrightarrow[]{l \to \infty}  -g_{s}^{2} e C_{F} \left [ \frac{\slashed{l} \gamma^{\mu} \slashed{l}}{(l^{2} - M^{2})^{3}}\right] . 
\end{align}
The UV limit of the $\Delta^{\mu}_{\gamma}(r,q,k,l)$ counterterm vertex in Eq.~\eqref{eq:DeltaTransient} is given by, 
\begin{align}
    \Delta^{\mu}_{\gamma}(r,q,k,l) \xrightarrow[]{l \to \infty} \Delta^{\mu}_{\gamma, UV}(r,q,k,l) = g_{s}^{2} e C_{F} (2-D) \frac{\slashed{\eta}}{q \cdot \eta} \, \left [\frac{2 l^{\mu}(l \cdot q)}{(l^{2}-M^{2})^{3}} \right] . 
\end{align}
After a tensor reduction, $\Delta^{\mu}_{\gamma, UV}(r,q,k,l)$ vanishes upon integration as the contraction with the photon polarization generates the term  $q \cdot \epsilon^{*} = 0 $. It exactly cancels the transient singularity of the UV counterterm in Eq.~\eqref{eq:abelian-vertex-UV2}, 
\begin{align}
    -g_{s}^{2} e C_{F} \frac{i}{\slashed{r}-\slashed{q}} \left \{ \left[\frac{\slashed{l} \gamma^{\mu} \slashed{l}}{(l^{2} - M^{2})^{3}} \right ] - \left[ \frac{\slashed{\eta}}{q \cdot \eta} \, \frac{2 l^{\mu}(l \cdot q)}{(l^{2}-M^{2})^{3}} \right] \right \} \frac{i}{\slashed{r}} \, \epsilon^{*}_{\mu} \xrightarrow{r = xq} 0 . 
\end{align}
\section{Outlook \& Conclusion}
In these proceedings, we have presented the construction of ultraviolet counterterms essential for a locally finite amplitude for photoproduction via quark annihilation, complementing the treatment of transient infrared singularities in Ref.\,\cite{Anastasiou:2025cvy}.  
We have demonstrated that these UV subtractions satisfy the necessary Ward identity cancellations, which ensures the local factorisation of initial-state infrared singularities in all soft and collinear limits while simultaneously eliminating transient final-state singularities. Thus, our two-loop locally finite amplitude integrand for general colorless production is ready for numerical integration in loop momentum space in four spacetime dimensions using methods developed in Refs.\,\cite{Kermanschah:2025wlo,Kermanschah:2024utt,Soper:1999xk,Nagy:2006xy,Gong:2008ww,Becker:2010ng,Assadsolimani:2009cz,Becker:2012aqa,Becker:2011vg,Becker:2012bi,Gnendiger:2017pys,Seth:2016hmv,Capatti:2019edf,Capatti:2020ytd,Capatti:2020xjc,TorresBobadilla:2020ekr,Kermanschah:2021wbk,Rios-Sanchez:2024xtv}. This work provides the foundation for extending our framework to processes involving jet production. 

\acknowledgments
We would like to thank the conference organisers of RADCOR 2025 for their kind hospitality. 
This work was supported in part by the National Science Foundation, award PHY-2210533, and the Swiss National Science Foundation, grant 10001706.
\bibliographystyle{JHEP}
\bibliography{biblio.bib}

\end{document}